\documentclass[journal=jpccck,manuscript=article]{achemso}

\usepackage{chemformula} 
\usepackage[T1]{fontenc} 

\author{Petr Liška}
\affiliation{Institute of Physical Engineering, Faculty of Mechanical Engineering, Brno University of Technology, Technická 2896/2, 616 69 Brno, Czech Republic}
\alsoaffiliation{Central European Institute of Technology, Brno University of Technology, Purkyňova 123, 612 00 Brno, Czech Republic}
\email{petr.liska1@vutbr.cz}

\author{Tomáš Musálek}
\affiliation{Institute of Physical Engineering, Faculty of Mechanical Engineering, Brno University of Technology, Technická 2896/2, 616 69 Brno, Czech Republic}

\author{Tomáš Šamořil}
\affiliation{Institute of Physical Engineering, Faculty of Mechanical Engineering, Brno University of Technology, Technická 2896/2, 616 69 Brno, Czech Republic}
\alsoaffiliation{Central European Institute of Technology, Brno University of Technology, Purkyňova 123, 612 00 Brno, Czech Republic}
\altaffiliation{TESCAN ORSAY HOLDING, a.s, Libušina tř. 21, Brno 623 00, Czech Republic}

\author{Matouš Kratochvíl}
\affiliation{Faculty of Chemistry, Brno University of Technology, Purkyňova 464/118, 612 00 Brno, Czech Republic}

\author{Radovan Matula}
\affiliation{Institute of Physical Engineering, Faculty of Mechanical Engineering, Brno University of Technology, Technická 2896/2, 616 69 Brno, Czech Republic}

\author{Michal Horák}
\affiliation{Institute of Physical Engineering, Faculty of Mechanical Engineering, Brno University of Technology, Technická 2896/2, 616 69 Brno, Czech Republic}
\alsoaffiliation{Central European Institute of Technology, Brno University of Technology, Purkyňova 123, 612 00 Brno, Czech Republic}

\author{Matěj Nedvěd}
\affiliation{Institute of Physical Engineering, Faculty of Mechanical Engineering, Brno University of Technology, Technická 2896/2, 616 69 Brno, Czech Republic}

\author{Jakub Urban}
\affiliation{Institute of Physical Engineering, Faculty of Mechanical Engineering, Brno University of Technology, Technická 2896/2, 616 69 Brno, Czech Republic}

\author{Jakub Planer}
\affiliation{Central European Institute of Technology, Brno University of Technology, Purkyňova 123, 612 00 Brno, Czech Republic}

\author{Katarína Rovenská}
\affiliation{Central European Institute of Technology, Brno University of Technology, Purkyňova 123, 612 00 Brno, Czech Republic}

\author{Petr Dvořák}
\affiliation{Institute of Physical Engineering, Faculty of Mechanical Engineering, Brno University of Technology, Technická 2896/2, 616 69 Brno, Czech Republic}
\alsoaffiliation{Central European Institute of Technology, Brno University of Technology, Purkyňova 123, 612 00 Brno, Czech Republic}

\author{Miroslav Kolíbal}
\affiliation{Institute of Physical Engineering, Faculty of Mechanical Engineering, Brno University of Technology, Technická 2896/2, 616 69 Brno, Czech Republic}
\alsoaffiliation{Central European Institute of Technology, Brno University of Technology, Purkyňova 123, 612 00 Brno, Czech Republic}

\author{Vlastimil Křápek}
\affiliation{Institute of Physical Engineering, Faculty of Mechanical Engineering, Brno University of Technology, Technická 2896/2, 616 69 Brno, Czech Republic}
\alsoaffiliation{Central European Institute of Technology, Brno University of Technology, Purkyňova 123, 612 00 Brno, Czech Republic}

\author{Radek Kalousek}
\affiliation{Institute of Physical Engineering, Faculty of Mechanical Engineering, Brno University of Technology, Technická 2896/2, 616 69 Brno, Czech Republic}

\author{Tomáš Šikola}
\affiliation{Institute of Physical Engineering, Faculty of Mechanical Engineering, Brno University of Technology, Technická 2896/2, 616 69 Brno, Czech Republic}
\alsoaffiliation{Central European Institute of Technology, Brno University of Technology, Purkyňova 123, 612 00 Brno, Czech Republic}

\title[Correlative Imaging of Individual CsPbBr$_3$ Nanocrystals: Role of Isolated Grains in Photoluminescence of Perovskite Polycrystalline Thin Films]
  {Correlative Imaging of Individual CsPbBr$_3$ Nanocrystals: Role of Isolated Grains in Photoluminescence of Perovskite Polycrystalline Thin Films}

\abbreviations{LHP, PL, NC}

\begin{document}

\begin{tocentry}



\includegraphics{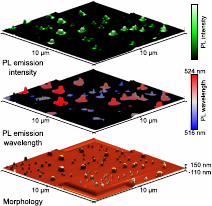}



\end{tocentry}

\begin{abstract}

We report on the optical properties of CsPbBr$_3$ polycrystalline thin film on a single grain level. A sample comprised of isolated nanocrystals (NCs) mimicking the properties of the polycrystalline thin film grains that can be individually probed by photoluminescence spectroscopy was prepared. These NCs were analyzed using correlative microscopy allowing the examination of structural, chemical, and optical properties from identical sites. Our results show that the stoichiometry of the CsPbBr$_3$ NCs is uniform and independent of the NCs' morphology. The photoluminescence (PL) peak emission wavelength is slightly dependent on the dimensions of NCs, with the blue shift up to 9\,nm for the smallest analyzed NCs. The magnitude of the blueshift is smaller than the emission linewidth, thus detectable only by high-resolution PL mapping. By comparing the emission energies obtained from the experiment and  a rigorous effective mass model we can fully attribute the observed variations to the size-dependent quantum confinement effect.
\end{abstract}

\section{Introduction}
Fully inorganic lead halide perovskite (LHP) CsPbBr$_3$ is a semiconducting material with a direct band gap exhibiting unique optical properties, such as high internal and external quantum yields of both Stokes and anti-Stokes photoluminescence (PL), narrow PL linewidth, low non-radiative losses, remarkable photostability, and high PL emission wavelength tunability due to a strong quantum confinement effect (QCE) \cite{Morozov20171013,Kovalenko20171109,Liu20171005,Tong20161024,VanLe2018,Raino2022,Pan2015,Akkerman2018,Wang20160113}. For its unique optical properties, CsPbBr$_3$ has emerged as a contemporary material that has fuelled further intensive development in the fields of optoelectronics, such as light-emitting devices \cite{Cheng2019,EZpYZ6JgWNnRNYHt}, single photon-emitters \cite{Zhu2022}, photovoltaics \cite{Jena20190313,Tong2020,Duan20210611}, high-energy gamma radiation detectors \cite{He2021}, photocatalysis of chemical processes \cite{Xu20170417}, high-resolution displays \cite{Ko20210310}, non-linear optical wavelength converters \cite{Qiu20180910}, reconfigurable memristors \cite{John2022}, hyper-sensitive scintillators \cite{Ma2021,Xu2020,Heo2018} or even functional metasurfaces \cite{Fan20210908}. The exceptionality of CsPbBr$_3$ originates from the unique electronic fine structure and high tolerance towards structural defects \cite{Becker2018, Tamarat2023}. Moreover, due to the inherent ionic character of CsPbBr$_3$, a cheap and facile fabrication of CsPbBr$_3$ nanocrystals (NCs) is possible simply by mixing the corresponding precursor solutions without the need for elevated temperature or other, potentially challenging conditions \cite{Zhang20150428}. Hence, NCs and their thin films in general form the basis of halide perovskite-based optoelectronic devices with unique or improved optical and electronic properties \cite{Xue2020}. The optical and electronic properties of CsPbBr$_3$ NCs are strongly dependent on their dimensions, morphology, size distribution and surface passivation \cite{Luo20190313,AbdiJalebi2018}. Decreasing the size of CsPbBr$_3$ crystals to the nanoscale or appropriate surface passivation can be utilized to tune or even enhance their optical properties \cite{Wei2019,Sichert20151014,Li20180621,Lin2021,Cheng20200625,Imran2016}. A full control over the size of the colloidal lead halide perovskite nanocrystals (cn-LHP) and its distribution has been demonstrated by Protesescu et al., who presented a hot-injection method allowing the colloidal synthesis of well-defined, monodisperse and monocrystalline NCs with PL properties far exceeding the polycrystalline films (pf-LHP) \cite{Protesescu20150610}. 

However, cn-CsPbBr$_3$ prepared by the hot-injection method suffer from a significant underperformance due to problems with electrical contacting which substantially limits their usage in optoelectronic devices in comparison to their polycrystalline film (pf-CsPbBr$_3$) counterparts \cite{Shamsi2021}. Therefore, the prospect of long operational devices stability, high conversion efficiency (>25\,\%), and simple utilization in electroluminescent devices make pf-CsPbBr$_3$ better candidates for today's optoelectronic devices than cn-CsPbBr$_3$, especially for solar cells technology \cite{Shamsi2021,Steele2021,Jiang2019,Yoshikawa2017}. 

A suitable way to study the optical properties of pf-CsPbBr$_3$ such as the emission wavelength, spectral broadening, recombination kinetics processes, and internal electrochemical potential of free charge carriers is through PL spectroscopy \cite{Morozov20171013,Kirchartz2020,DiStasio20170926,Zhou20210315,Dey2018,Zhang2015}. When correlated with the high-resolution imaging techniques of low-dimensional structures, PL spectroscopy provides a direct insight into the influence of local morphology effects \cite{Anni2021} or the energy levels shift caused by QCE \cite{Butkus2017,Luo2019,Cheng2020}. However, due to the limited spatial resolution of optical microscopy, this approach cannot provide a response of individual grains of polycrystalline thin film, as there are usually several grains within the focus.    

Here, we demonstrate a comprehensive analysis of the material properties of individual CsPbBr$_3$ NCs prepared from a low-concentrated pf-LHPs solution while achieving considerable spacing between individual NCs. This system is representative for thin polycrystalline films, as it was grown under the conditions typically used for the growth of thin films. In contrast to the thin polycrystalline film with overlapping grains, the well-isolated NCs can be addressed individually by optical spectroscopy. A correlative approach based on focused ion beam (FIB) tagging of the examined area on the sample is used and the synergy of high-resolution experimental techniques is utilized: transmission electron microscopy (TEM), scanning electron microscopy (SEM), atomic force microscopy (AFM) and confocal optical spectroscopy (COS) together with anti-Stokes PL mapping. These techniques are capable of analysing CsPbBr$_3$ NCs' inner structure, determining their characteristic dimensions, morphology, and their PL response. We experimentally retrieve the dependence of the PL peak emission wavelength on the characteristic volume and aspect ratio of the NC at a single NC level. We demonstrate that the emission energy is governed by a size-dependent QCE predictable within a simple effective mass model. The understanding of the relation between the characteristic dimension of individual CsPbBr$_3$ NCs and their emission energy is a crucial element in their implementation in advanced optoelectronic devices. 

\section{Methods}
\subsection{Fabrication of CsPbBr$_3$ NCs}
Transparent fused silica substrates covered by an indium-tin-oxide (ITO) layer with a thickness of 50\,nm were subsequently cleaned by acetone (5 min.), isopropylalcohol (5 min.) and deionized water (5 min.) baths in ultrasound. After the deionized water treatment, the residues of water and dust particles were blown off with a nitrogen and by placing the substrates on a hot-plate (100\,$^\circ$C) for 10 minutes. For the synthesis of CsPbBr$_3$ NCs, a saturated solution of CsBr and PbBr$_2$ precursors dissolved in dimethylformamide (DMF) was dropcasted onto the substrates and left dry out at room temperature.

\subsection{Correlative imaging of CsPbBr$_3$ NCs}

The FIB micro-markings for navigation across the sample and correlative imaging were fabricated using a focused ion beam/scanning electron microscope TESCAN LYRA3. The energy of Ga$^+$ primary ions was 30\,keV, ion beam current 2\,nA, and the target depth of the micro-markings was an equivalent of 200\,nm in a Si substrate. 

The morphology of the CsPbBr$_3$ NCs was obtained by subsequent analysis by SEM and AFM. The SEM measurements were performed using high-resolution SEM FEI Verios 460L at 3\,kV and 13\,pA. The AFM measurements were performed using a scanning probe microscope Bruker Dimension Icon with ScanAsyst-Air high-resolution imaging probes with triangular geometry and a tip radius of 12\,nm.

The optical properties of individual CsPbBr$_3$ NCs were measured by PL spectroscopy. The PL maps were obtained with Witec Alpha 300R confocal microscopy and optical spectroscopy system. The laser light with excitation wavelength of 532\,nm and linewidth <1\,nm was used. The observed PL spectra are of single-photon phonon-assisted anti-Stokes PL origin and nearly identical to the PL spectra with Stokes origin. As it has been demonstrated, both the Stokes and anti-Stokes excitation yield nearly identical PL spectra \cite{Morozov2017}. The obtained PL maps have a resolution of 120$\times$120\,pixels$^2$ and they were obtained with an objective Zeiss EC Epiplan-Neofluar Dic with the 100$\times$ magnification and numerical aperture of 0.9 and 600\,grating/mm diffraction grid. The spectra were collected by utilizing the laser light with optical power of 37\,\textmu W, and integration time of 0.1\,s. The parameters $I_0$ -- PL peak intensity, $\lambda_0$ -- PL peak emission wavelength, and $\lambda_\mathrm{FWHM}$ -- PL full width-at-half-maximum (FWHM) were obtained by fitting the Gaussian function to the background-subtracted PL spectra

\begin{equation}
I(\lambda)=I_0\exp{\left\{-\frac{1}{2}\left(\frac{\lambda-\lambda_0}{\lambda_\mathrm{FWHM}}\right)^2\right\}}.
\end{equation}

The chemical imaging was done by time-of-flight secondary ion mass spectrometry (ToF-SIMS) and x-ray photoelectron spectroscopy (XPS) instrumentations. The ToF-SIMS measurements were performed using a TESCAN AMBER focused ion beam - scanning electron microscope (FIB-SEM) equipped with an orthogonal ToF-SIMS system (C-TOF module provided by TOFWERK). The energy of Ga$^+$ primary ions was 30\,keV, ion beam current 47\,pA, and pixel dwell time 11\,\textmu s. The Br 3D chemical image was done in the negative ion mode and Ga, Cs and Pb maps were obtained in the positive ion mode. The XPS measurements were performed using an x-ray photoelectron spectroscopy axis supra (KRATOS-XPS). The XPS spectra were fitted using Lorenz curve and U2 Tougaard background subtraction. The orbitals used for determining the stoichiometry were: Cs 3d$_{3/2}$, Cs 3d$_{5/2}$, Pb 4f$_{5/2}$, Pb 4f$_{7/2}$, Br 3d$_{3/2}$, and Br 3d$_{5/2}$.

The structural analysis of CsPbBr$_3$ NCs was carried out by high-resolution TEM analysis. TEM, STEM and electron energy loss spectroscopy (EELS) measurements were conducted by a (Scanning) Transmission Electron Microscope FEI Titan Themis. High-resolution imaging was performed in TEM mode at 300\,keV.  STEM imaging and EELS were performed in a monochromated scanning regime at 120\,keV while the convergence semi-angle was set to 10\,mrad and the collection semi-angle for EELS to 14.4\,mrad.

\subsection{Density functional and effective mass theory}
As a starting point for the CsPbBr$_3$ elementary cell volume relaxation, which enabled the theoretical computation of the CsPbBr$_3$ electronic structure and further structural analysis, the experimentally obtained lattice parameter $a=5.9$\,\AA\,\, was used. The calculation at fixed volume utilized the lattice parameter which varied within $a\pm7$\,\%.  By conducting volume relaxation of the elementary cell, it is possible to retrieve the dependence of the free energy per unit cell $F$ on the volume of the CsPbBr$_3$ elementary cell $V$. The minimum of the $F(V)$ function corresponds to the equilibrium elementary cell volume $V_0$ and thus to the equilibrium lattice constant $a_0$, obtained by fitting $F(V)$ by the Murnaghan equation of state 

\begin{equation}
P(V)=\frac{B_0}{B_0'}\left[\left(\frac{V_0}{V}\right)^{B_0'}-1\right]
\end{equation}

\noindent where $P$ is the pressure acting on the unit cell acquired by the partial derivation $P(V)=-\left(\partial F/\partial V\right)_{\textmd{T,N}}$ at a constant temperature $T$ and with a constant number of particles in the enclosed system $N$. The calculated lattice parameter was utilized in the density functional theory (DFT) calculations of the electronic structure of CsPbBr$_3$.

All DFT calculations were performed with the Vienna ab initio Simulation Package (VASP) \cite{Kresse1993, Kresse1994, Kresse1996, Kresse1996-2, Kresse1999, Becke19930115}, using the projector-augmented wave method \cite{Kresse1999} for treating core electrons. The Bloch functions for nine valence electrons of caesium (5s$^2$5p$^6$6s$^1$), fourteen valence electrons of lead (6s$^2$5d$^{10}$6p$^2$), and seven valence electrons of bromine (2s$^2$5p$^5$), were expanded in a plane wave basis set with an energy cut-off 400\,eV. The Brillouin zone was sampled with a Gamma-centered $6\times6\times6$ Monkhorst-Pack grid \cite{Pack1977}. The self-consistent electronic calculations converged to 10$^{-5}$\,eV. Spin-orbit coupling was taken into account in all calculations. We used the PBEsol functional \cite{Perdew2008} for geometry optimization of the cubic CsPbBr$_3$. The electronic band structure of the cubic CsPbBr$_3$ was calculated with a modified HSE06 functional \cite{Krukau2006} with $45\,\%$ Hartree-Fock mixing. The graph of the electron structure and the values of effective masses were retrieved with the help of the program Vaspkit \cite{VASPKIT}. The effective masses were obtained by a third-order polynomial fit of the energy bands. 

The electronic band gap of CsPbBr$_3$ $E_\textmd{g}=2.425$\,eV used in the effective mass theory calculations was determined by fitting the experimental data with the theoretical model (eq.\,4). The binding energy of an exciton in CsPbBr$_3$ was determined by the hydrogen atom model

\begin{equation}
    E_\textmd{b}=\frac{m^*e^4}{8h^2\varepsilon^2\varepsilon_0^2}=47\,\textmd{meV},
\end{equation}

\noindent where $m^*=0.08m_\textmd{e}$ and $\varepsilon=4.8$ is the experimental permittivity value for CsPbBr$_3$ \cite{Becker2018}.

\section{Results and discussion}

\begin{figure}[hbtp!]
\centering
\includegraphics[width=\textwidth]{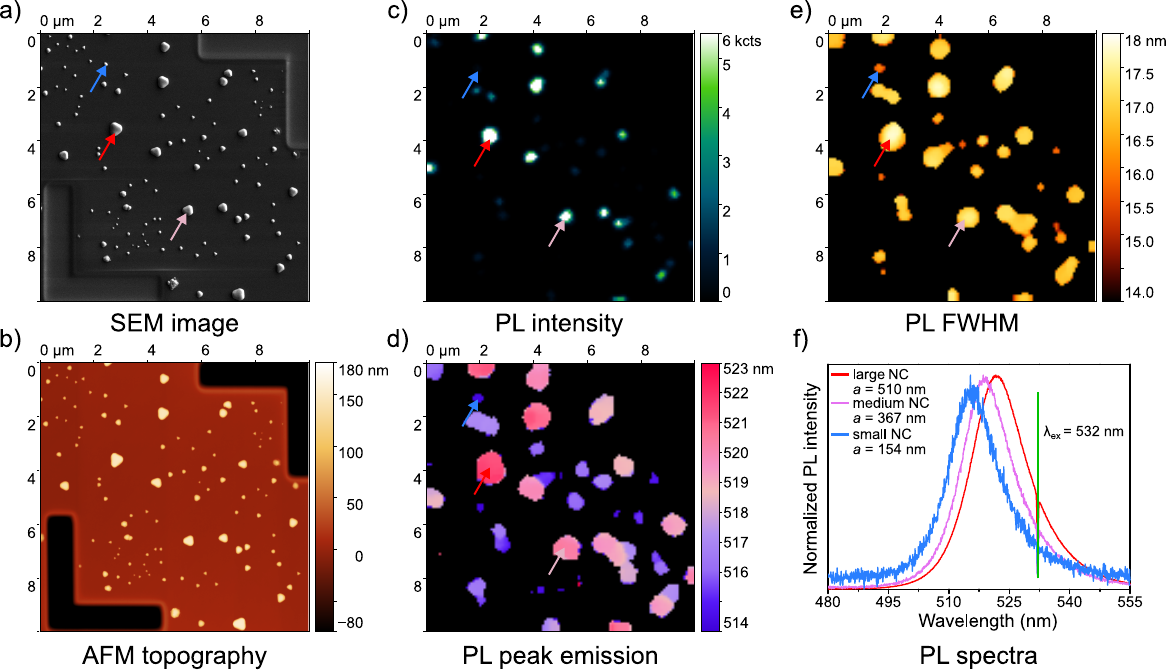}
\caption{a) SEM image showing the CsPbBr$_3$ NCs of various sizes and shapes. b) AFM topography image of the identical area displaying the height profile of the NCs. c) PL intensity, d) PL peak emission wavelength and e) PL FWHM maps acquired by measuring PL spectra in the examined area and regression analysis of the background--corrected PL spectra. f) Representative PL spectra of three CsPbBr$_3$ NCs marked by the arrows in the SEM image and PL maps. The characteristic widths \textit{a} are indicated for each of the NCs.}
\label{fig1}
\end{figure}

Figure~\ref{fig1}a shows an SEM image of the sample with CsPbBr$_3$ NCs and and FIB micro-markings used for the navigation across the sample and correlation of the obtained data. Here, we observe NCs of various shapes and sizes with the size distribution ranging from tens to hundreds of nm with the most frequent values peaking around 120\,nm. The SEM image is accompanied by the AFM topography image (Figure~\ref{fig1}b) of the identical sample area which displays the height distribution of the NCs ranging from 10$^1$ to 10$^2$\,nm. Both SEM and AFM images are utilized to determine the aspect ratio $AR=a/c$ and the volume $V$ of NCs, based on the characteristic width $a$ and height $c$ which are measured by SEM and AFM, respectively. 

The morphology is correlated with PL maps in Figures~\ref{fig1}c--e. The PL maps were obtained by COS -- every pixel of the PL map has an assigned PL spectrum -- and subsequent regression analysis of the PL spectra (see Methods). In Figure~\ref{fig1}c we see the PL intensity map which matches the positions of the CsPbBr$_3$ NCs visible in Figures~\ref{fig1}a and b. By comparing these images, it is evident that bigger NCs exhibit brighter PL. Figure~\ref{fig1}d displays the PL peak emission wavelength of the NCs ranging from 514\,nm (2.41\,eV) for the smaller NCs to 523\,nm (2.37\,eV) for the bigger NCs. The PL FWHM map (Figure~\ref{fig1}e) shows the values ranging from 14\,nm (68\,meV) for smaller NCs to 18\,nm (80\,meV) for the bigger ones. Thanks to the correlation of structural Figure~\ref{fig1}a,b and optical maps Figure~\ref{fig1}c,d,e, we were able to assign particular PL spectra to individual CsPbBr$_3$ NCs (Figure~\ref{fig1}f). 

\begin{figure}[hbtp!]
\centering
\includegraphics[width=0.75\textwidth]{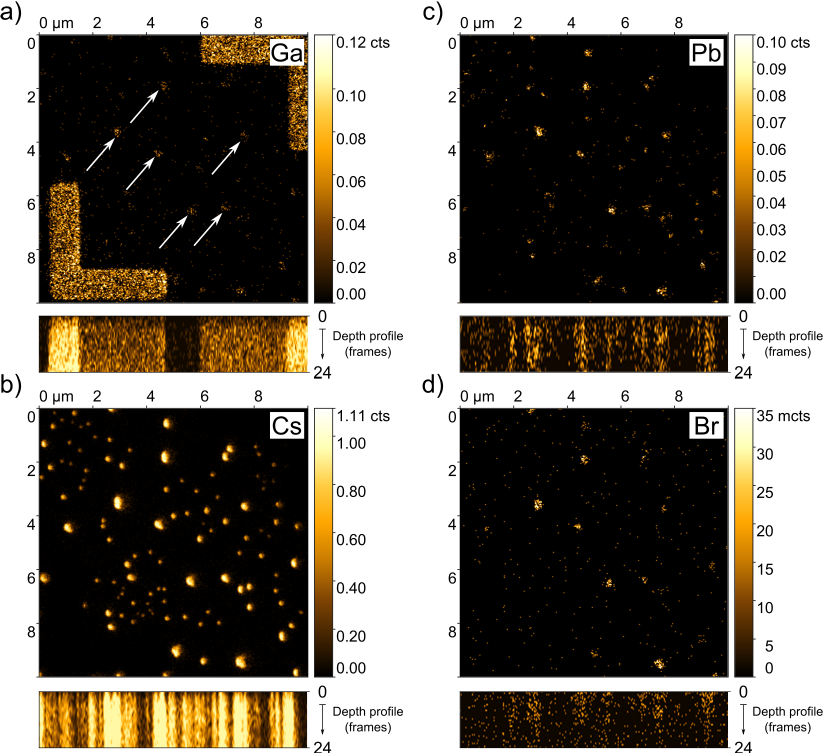}
\caption{a) Analysed region on the sample used for the 3D elemental mapping of Ga, b) Cs, c) Pb and d) Br atoms. The white arrows in a) show the Ga ion implementation into the CsPbBr$_3$ NCs. The intensity of the signal obtained by ToF-SIMS in b)--d) does not indicate any inhomogeneity in the spatial or depth elemental distribution. This assumption is based on a comparison of ToF-SIMS signals between NCs with similar volumes.}
\label{fig2}
\end{figure}

To exclude the role of local stoichiometry on the variance of PL properties of individual NCs, correlative elemental imaging by ToF-SIMS was performed. In Figure \ref{fig2}a--d, an analysed area is shown (the same as in Figure~\ref{fig1}) to which are correlated the 3D spatial distributions of the detectable elements: Ga, Cs, Pb and Br ions acquired via ToF-SIMS analysis. Since Ga ion beam was used for both FIB micro-marking and ToF-SIMS elemental imaging, it is expected that the Ga ions will be present on the sample surface. From Figure \ref{fig2}a it seems, that some Ga ions were implanted also inside CsPbBr$_3$ NCs (see the white arrows in Figure \ref{fig2}a). The fabrication of FIB micro-markings could in principle lead to the alteration of structural and optical properties of the examined samples. The influence of Ga ions implanted during the fabrication of FIB micro-markings on the optical properties of CsPbBr$_3$ NCs was studied as follows. The PL response of several NCs was measured before and after their exposure to gallium ions used for the fabrication of the FIB markings. We discovered that the exposure to Ga ions of applied dose decreases the PL integral intensity by about 6\,\% but does not alter the PL peak emission wavelength or FWHM (see Supporting information, section S2). Figures~\ref{fig2}b--d demonstrate the identical distribution of Cs, Pb, and Br elements throughout the whole volume of the sample. Further, integral XPS analysis points to high chemical purity of the NCs with stoichiometry Cs$_{1.2}$Pb$_1$Br$_{2.8}$ (see Supporting information, section S3). These indicate that variations observed in the PL peak emission wavelength might be rather attributed to the morphology differences of the individual NCs and the related QCE.

\begin{figure}[hbtp!]
\centering
\includegraphics[width=\textwidth]{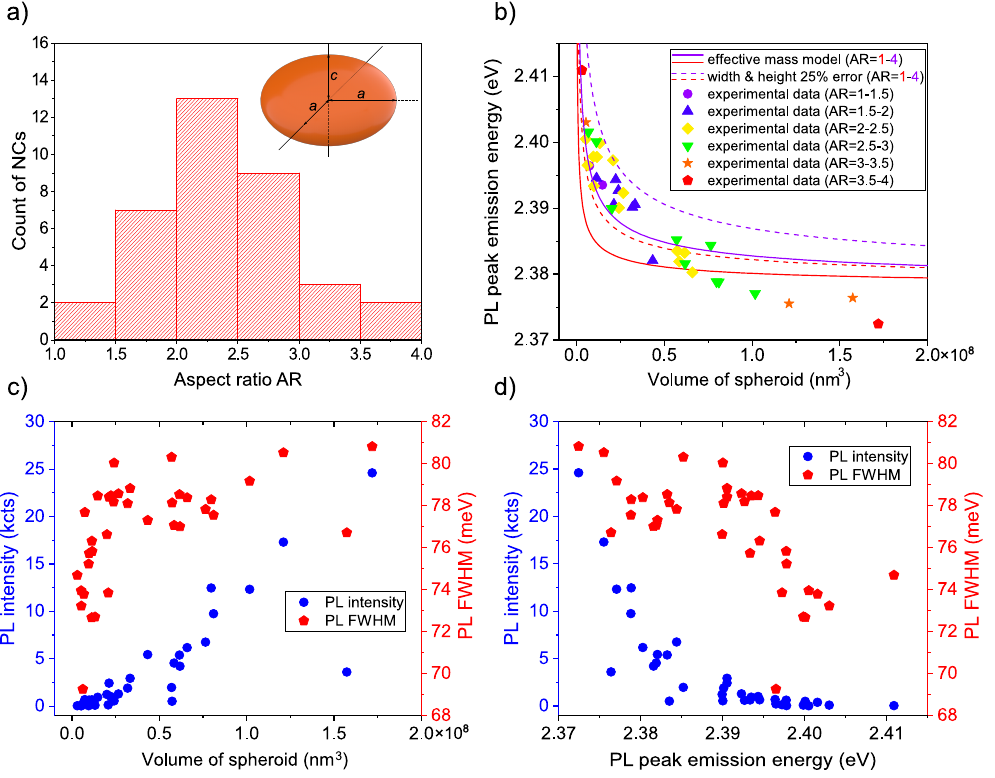}
\caption{a) Histogram of the NCs aspect ratio $a/c$ ranging in between 1 and 4 with the 3D model of a spheroid used to represent the NCs by the main axis parameters $a$ and $c$. b) The experimentally obtained PL peak emission energies of individual CsPbBr$_3$ NCs of particular aspect ratios are plotted as a function of the spheroid volume and compared to the QCE models predicting the volume-and-aspect-ratio-dependent PL peak energy shift. The dashed lines correspond to the assumption that the effective dimensions of NCs are by 25\,\% smaller than those determined from SEM and AFM images. The PL intensity and PL FWHM have been plotted as the functions of c) volume of the NCs and d) PL peak emission energy.}
\label{fig3}
\end{figure}

To determine the dependency of the CsPbBr$_3$ NCs' PL response on their morphology, we have applied the quantum confinement effect (QCE) model based on the effective mass theory and parameters obtained by density functional theory (DFT) (see Methods and Supporting information, section S1). For the purpose of the simulations, we have proposed a model shape of NCs as a spheroid with only two parameters: the volume and the aspect ratio. Even though this shape does not fully correspond to the real image of the CsPbBr$_3$ NCs, it respects the real aspect ratios. We were able to obtain the aspect ratio $AR=a/c$, where $a$ is the length of the NCs (Figure~\ref{fig1}a) and $c$ is the height of the NCs (Figure~\ref{fig1}b) from the CsPbBr$_3$ NCs' morphology. In Figure~\ref{fig3}a, the histogram of the aspect ratios of individual NCs with values ranging from 1 to 4 is shown, with the values $2-2.5$ being the most common. This means that the lateral dimensions of the CsPbBr$_3$ NCs tend to be larger than the vertical dimensions. The total volume of a such spheroid is then calculated as $V=\frac{4}{3}\pi a^2c$.

Let us consider an infinite potential well with the shape of a spheroid representing a single CsPbBr$_3$ NC. The energy level distribution for an exciton confined within this system can be expressed as 

\begin{equation}
    E_{smn}=(E_\textmd{g}-E_\textmd{b})+\frac{\hbar^2}{2m^*}\left[\frac{\tau_{s,m}^2}{a^2}+\frac{2\tau_{s,m}}{ac}\left(n+\frac{1}{2}\right)\right], \hspace{5mm} \frac{1}{m^*}=\frac{1}{m_{\textmd{e}}^*}+\frac{1}{|m_{\textmd{h}}^*|},
    \label{eq1}
\end{equation}

\noindent where $E_\textmd{g}=2.425$\,eV is the band gap of CsPbBr$_3$, $m^*=0.08m_\textmd{e}$~is the reduced effective exciton mass (see Supporting information, section S1 for values of the parameters), $E_\textmd{b}=47$\,meV is the binding energy of an exciton accounting for the Coulomb interaction between an electron and a hole (see Methods), $\tau_{s,m}$ is the $s^\textmd{th}$ root of the Bessel function of the first kind where $s=(0,1,2,\dots)$, $m=(1,2,3,\dots)$, and $n=(0,1,2,\dots)$ as described in Ref.~\cite{Kereselidze2016}. The experimentally and theoretically retrieved values for the PL peak emission energy of individual NCs are plotted and compared as a function of a spheroid volume with a varying aspect ratio in Figure~\ref{fig3}b. The theoretical values are systematically somewhat larger then the experimental ones. A plausible explanation is that the effective dimensions of NCs are smaller than the dimensions determined from the SEM and AFM images. Indeed, the wave function needs to vary smoothly in space. In the realistic NCs of irregular shape, the wave function cannot exploit the full volume of the particle. Instead, it approximately takes a spheroidal shape inscribed to the NC (see e.g. Fig.~4 in Ref.~\cite{Huo2014}). Consequently, the effective dimensions of NCs are smaller. By analysis of the realistic particle shapes we estimate the difference of the effective and real dimension to be up to 25\,\% (see Supporting information S4). We show the corresponding theoretical dependences assuming the effective dimensions of NCs by the dashed lines in Figure~\ref{fig3}b. Now, most of the experimental points agree well with this corrected theoretical prediction. It is also notable, that the amplitude of the PL peak emission energy shift ($\approx39\,$meV$/9$\,nm) being smaller than the average PL FWHM ($\approx75$\,meV/$16$\,nm) results in a uniform macroscopic optical response of the hereby presented statistical ensemble of CsPbBr$_3$ NCs across the sample with the PL peak emission energy shift detected only by high-resolution optical imaging.

In Figure~\ref{fig3}c a comparison is made between the experimentally determined PL intensity, PL FWHM, and NCs' volume. Importantly, the PL intensity exhibits a linear dependence on the particle volume with 75\,\% of NCs having a deviation from the linear relationship bellow 20\,\% (blue points in Figure~\ref{fig3}c). The effect of non-radiative processes, which could have significantly decreased the intensity of small particles on the surface, is insignificant. Notably, only 11\,\% of the particles show a considerably lower PL intensity, which can be attributed to structural defects, lattice strain, presence of the trap states or other non-radiative recombination channels \cite{Jones2019, Wang2020}. Consequently, the total emission intensity is primarily linked to the volume of the material, indicating that the grain size does not have a significant impact. In Figure~\ref{fig3}d, a comparison is made between experimentally determined PL intensity, PL FWHM, and PL peak emission energy. The PL intensity significantly decreases with the increasing PL peak emission energy. The FWHM of the emission exhibits a slight but systematic increase for larger particles. At present, the origin of this effect is unknown to us. We note that for very small particles (several units of nm) the opposite trend has been observed previously and attributed to the size dependence of the phonon-exciton coupling strength \cite{Cheng20200625}. 

\section{Conclusions}

In conclusion, we have performed a comprehensive analysis of CsPbBr$_3$ NCs at an individual NC level. It consisted of correlative high-resolution morphological and optical analysis which has been used to determine the relation between the shape and dimensions of NCs and their PL peak emission energy. These experimental results have been used as input to a simplified QCE model based on the effective mass approach. We have been able to successfully predict the PL peak emission energy for CsPbBr$_3$ NCs based on the model of an exciton confined within the infinite spheroidal potential well. Our correlative analysis of a large statistical ensemble of individual CsPbBr$_3$ NCs has the following implications for their utilization in various optoelectronic applications: (i) CsPbBr$_3$ NCs produced by the simple drop-casting method are of high quality, exhibiting bright and size-tunable PL emission. (ii) The simplified and aspect ratio-based QCE model with effective masses obtained from DFT calculations qualitatively agrees with observed PL peak emission energies of CsPbBr$_3$ NCs without the need for complex numerical simulations. Despite relatively big variations in the shape and size of individual NCs, the observed QCE is moderate, up to 9\,nm in the wavelength. This is considerably smaller than their average PL FWHM, thus being detectable only by high-resolution PL mapping. (iii) The PL intensity exhibits a linear dependence on the particle volume and thus the total PL emission intensity of the NCs is primarily linked to the volume of the material, indicating that the grain size does not have a significant impact. (iv) CsPbBr$_3$ NCs are compatible with the FIB processing with Ga ions, which poses an opportunity for precise nanoprocessing of individual NCs, as well as their integration into more complex devices including optical cavities or waveguides.

\begin{acknowledgement}

The authors acknowledge support by Czech Science Foundation, grant No. 19-06621S, Ministry of Education, Youth, and Sports of the Czech Republic (projects CzechNanoLab, No. LM2018110, Quality Internal Grants of BUT, No.\,CZ.02.2.69/0.0/0.0/19\underline{\,\,\,}073/0016948, sub-grant CEITEC VUT-K-22-7680), and Specific Research of BUT FCH/FSI-J-21-7396. The authors acknowledge Prof. Maksym Kovalenko's laboratory for providing some perovskite mixtures used in this work.

\end{acknowledgement}

\begin{suppinfo}

The Supporting information is available.

\end{suppinfo}

\bibliography{achemso-demo}

\end{document}


\subsection{S1: Electronic and structural properties of CsPbBr$_3$ NCs}
To predict the influence of the CsPbBr$_3$ NCs' morphology on their PL properties, we have built a quantum confinement model based upon high-resolution TEM measurements and effective mass theory with parameters obtained by DFT. Figure~\ref{sfig1}a displays atomic-resolution TEM images of the CsPbBr$_3$ NCs (see Supporting Info S2), measured in order to obtain the lattice parameter of the NCs. The atomic-resolution images of CsPbBr$_3$ NCs were then processed by 2D fast Fourier transform (2D FFT), shown in Figure~\ref{sfig1}b in order to determine the lattice constant. To determine the experimental value of the CsPbBr$_3$ band gap, electron energy loss spectroscopy (EELS) measurements took place (Figure~\ref{sfig1}c), which indicated the band gap of CsPbBr$_3$ to be $E_\textmd{g,e}=(2.3\pm 0.1)$\,eV. The experimentally obtained lattice parameters from several NCs were displayed in a histogram (Figure~\ref{sfig1}d) which was fitted by a normal distribution. The statistically determined value of the lattice parameter was determined to be $\tilde{a}=(5.9\pm0.1)$\,\AA\,\,which is in good agreement with the value reported in the literature $a=5.865$\,\AA \cite{Mehl2017}. 

Figure~\ref{sfig1}e displays DFT calculated free energies as the function of the unit cell volume, and a fit to the Murnaghan equation of states (see Methods). The equilibrium lattice constant of CsPbBr$_3$ was found to be $a_0=5.864$\,\AA, which agrees well with the experiment. Figure~\ref{sfig1}f displays the band structure of bulk CsPbBr$_3$ calculated by DFT (see Methods) with equilibrium lattice constant as an input. Due to the nature of the DFT calculations, which artificially reduce the value of the band gap, the theoretical value of band gap was falsely determined as $E_\textmd{g,t}=1.48$\,eV. For the purpose of the QCE model in this paper, we have used the measured value of the band gap by EELS $E_\textmd{g,e}=(2.3\pm 0.1)$\,eV as a starting point for our simulations. By fitting the experimental data of the PL peak emission of individual NCs by eq.\,(4) in the main manuscript, the band gap was determined as $E_\textmd{g}=(2.425\pm0.002)$\,eV.  From the band structure, the electron and hole effective masses $m_\textmd{e}^*=0.167$\,$m_\textmd{e}$, $m_\textmd{h}^*=-0.153$\,$m_\textmd{e}$ in the direction $\Gamma$--R were obtained, where $m_\textmd{e}$ is the mass of a free electron (see Methods). These values are in good agreement with Ref.~\cite{Becker2018}, however, the incorrect value of the band gap suggests that these effective masses can vary by order of several percent.

\renewcommand{\thefigure}{S1}
\begin{figure}[hbtp!]
\centering
\includegraphics[width=\textwidth]{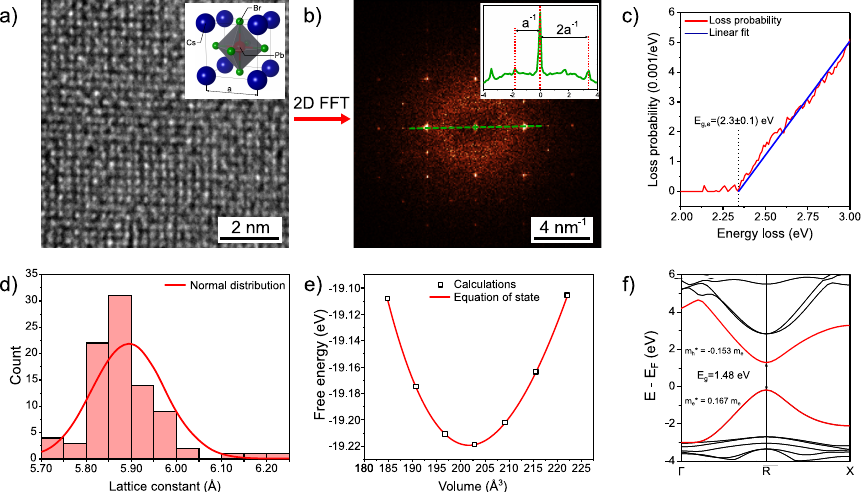}
\caption{a) Atomic-resolution TEM image of the analysed CsPbBr$_3$ NC with the model of the CsPbBr$_3$ cubic lattice. The high-resolution images post-processed in programme Gwyddion  \cite{Necas2012} via 2D-FFT to acquire the experimental lattice constant $a$. c) EELS loss probability spectrum of the CsPbBr$_3$ NCs with band gap determined experimentally as $E_\textmd{g,e}=(2.3\pm 0.1)$\,eV.
d) A histogram of lattice constants determined for 53 distinct CsPbBr$_3$ NCs fitted by normal distribution with the value of experimental lattice parameter $a=(5.9\pm0.1)$\,\AA. e) Volume relaxation calculations utilizing the experimental lattice parameter $a$ as a starting point, fitted by Murnaghan equation of state with an outcome of theoretical equilibrium lattice parameter $a_0=5.86$\,\AA. f) The theoretically calculated band structure of bulk crystal CsPbBr$_3$ acquired via DFT with an input parameter of theoretical equilibrium lattice parameter and output parameters of electron and hole effective masses for effective mass modelling of CsPbBr$_3$ NCs' PL response.}
\label{sfig1}
\end{figure}

\subsection{S2: Sensitivity of CsPbBr$_3$ nanocrystals to FIB and electron beam in TEM}

The exposure to the required dose of Ga ions during the fabrication of FIB micro-markings decreases the PL integral intensity by about 6\,\% (Figure~\ref{sfig2}) but does not alter the PL peak emission wavelength or FWHM.

\renewcommand{\thefigure}{S2}
\begin{figure}[hbtp!]
\centering
\includegraphics[width=0.55\textwidth]{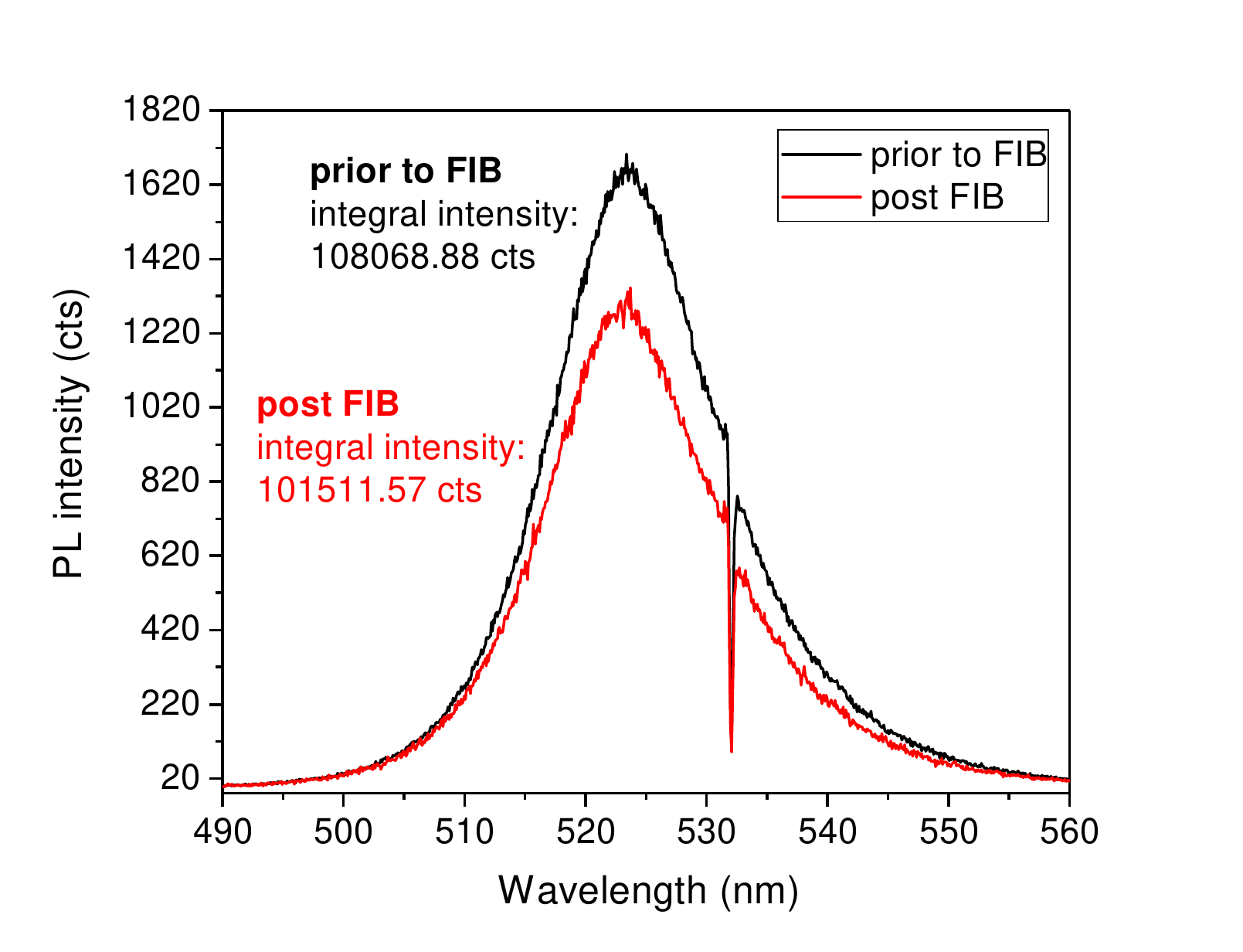}
\caption{Prior to FIB processing and post FIB processing obtained typical PL integral intensities of CsPbBr$_3$ NC.}
\label{sfig2}
\end{figure}

Due to the intense destruction of CsPbBr$_3$ NCs during the TEM analysis (Figure~\ref{sfig3}a), which resulted in the incapability of atomic-resolution images nor diffraction patterns, a compromise needed to be made. The acquisition of the TEM and EELS measurements has been made on the colloidal CsPbBr$_3$ NCs, which should exhibit the same crystalline structure as the CsPbBr$_3$ NCs prepared by drop casting. The STEM overview image of the colloidal nanocrystals is displayed in Figure~\ref{sfig3}b. The challenges of the CsPbBr$_3$ NCs and thin films imaging have been already discussed in \cite{Brennan2017,Duong2022} which results are in agreement with our observations.  
\renewcommand{\thefigure}{S3}
\begin{figure}[hbtp!]
\centering
\includegraphics[width=1\textwidth]{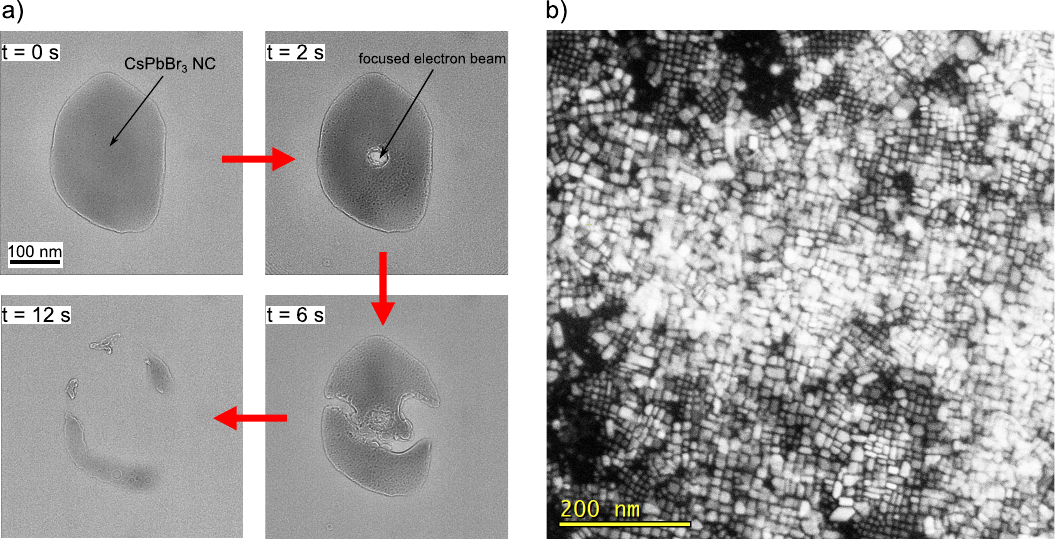}
\caption{a) Immediate degradation of drop casted CsPbBr$_3$ NCs in TEM observation under focused electron beam. b) Overview STEM image of colloidal CsPbBr$_3$ NCs serving for the crystal lattice acquisition.}
\label{sfig3}
\end{figure}

\subsection{S3: Stoichiometry of CsPbBr$_3$ nanocrystals}
To determine the purity of the prepared CsPbBr$_3$ NCs, integral XPS was done on the ensemble of the characterized NCs (Figure \ref{sfig4}). The resulting stoichiometry was determined as Cs$_{1.2}$Pb$_{1}$Br$_{2.8}$. 
\renewcommand{\thefigure}{S4}
\begin{figure}[hbtp!]
\centering
\includegraphics[width=0.9\textwidth]{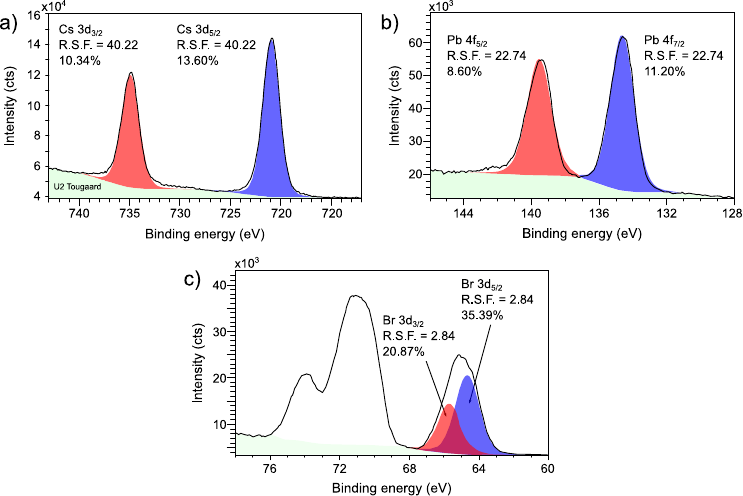}
\caption{XPS spectra for determination of the stoichiometry of a) Cs, b) Pb, and c) Br atoms.}
\label{sfig4}
\end{figure}

\subsection{S4: Error of the spheroidal approximation}
In order to estimate the error of the used spheroid approximation of the CsPbBr$_3$ NCs, the effective length $a$ and height $c$ of the NC have been compared to the lengths and heights of the inscribed $a_\textmd{in}, c_\textmd{in}$ (Figure~\ref{sfig5}a) and circumscribed $a_\textmd{out}, c_\textmd{out}$ (Figure~\ref{sfig5}b) spheroid. The effective lengths $a$ have been acquired from each NC by averaging over 12\,px from the horizontal profile in SEM and effective heights $c$ have been acquired by averaging over 12\,px from the topography of individual NC in the program Gwyddion \cite{Necas2012}. The lengths and heights of the inscribed or circumscribed spheroid were obtained by fitting the spheroid inside the NC or fitting the spheroid so it would contain the whole NC respectively. Figure~\ref{sfig5} displays the comparison between the a) inscribed or b) circumscribed lengths and heights of the NCs with the effective values. The average deviation of the used values and the effective ones was determined ranging between $5-25$\,\%. 
  
\renewcommand{\thefigure}{S5}
\begin{figure}[hbtp!]
\centering
\includegraphics[width=0.60\textwidth]{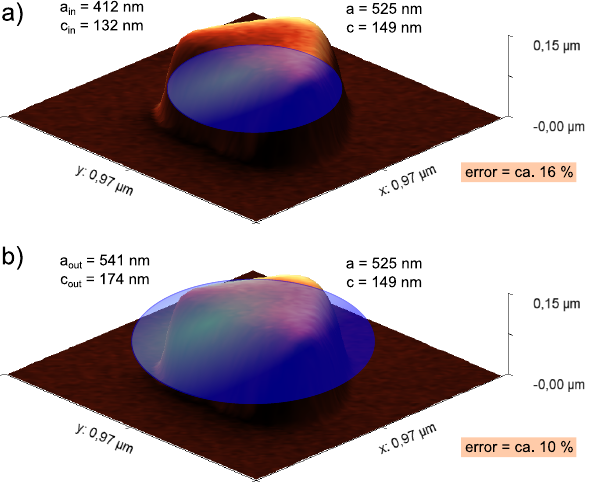}
\caption{a) Inscribed and b) circumscribed spheroid compared to the effective length and height of the NC.}
\label{sfig5}
\end{figure}

\subsection{S5: Optical properties of CsPbBr3 NCs in spectroscopic units}

For purposes of optical spectroscopy, we include Figure $3$ from the main manuscript also in the units of wavelengths (Figure \ref{sfig6}).

\renewcommand{\thefigure}{S6}
\begin{figure}[hbtp!]
\centering
\includegraphics[width=\textwidth]{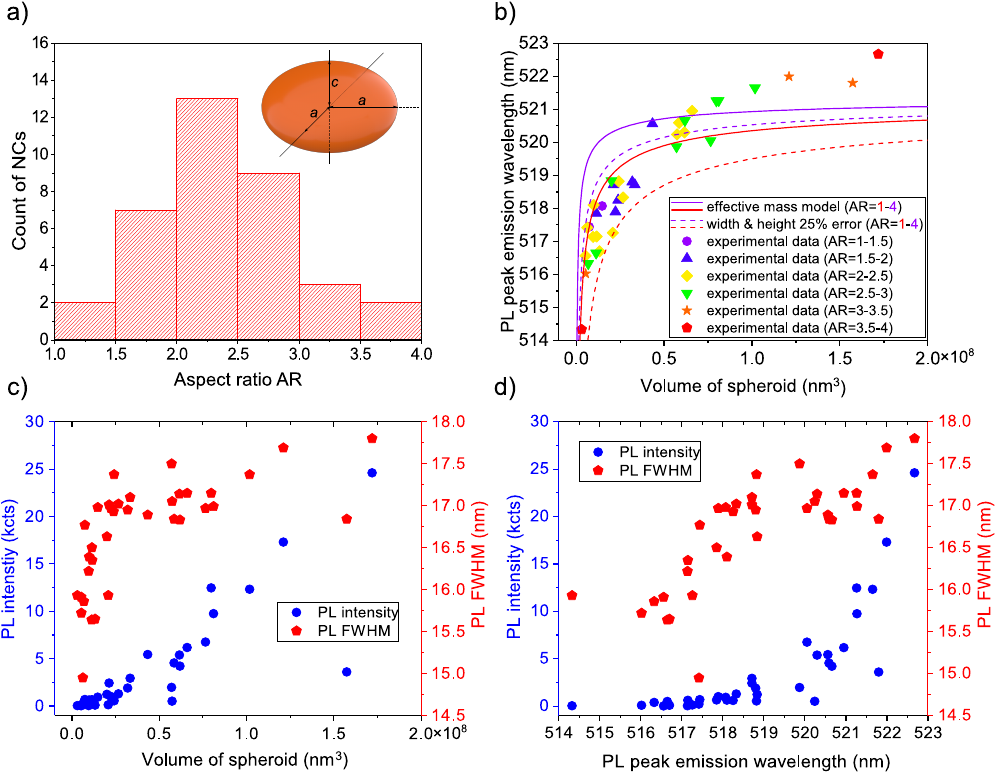}
\caption{a) Histogram of the NCs aspect ratio $a/c$ ranging in between 1 and 4 with the 3D model of a spheroid used to represent the NCs by the main axis parameters $a$ and $c$. b) The experimentally obtained PL peak emission wavelengths of individual CsPbBr$_3$ NCs of particular aspect ratios are plotted as a function of the spheroid volume and compared to the QCE models predicting the volume-and-aspect-ratio-dependent PL peak wavelength. The dashed lines correspond to the assumption that the effective dimensions of NCs are by 25\,\% smaller than those determined from SEM and AFM images. The PL intensity and PL FWHM have been plotted as the functions of c) volume of the NCs and d) PL peak emission wavelength.}
\label{sfig6}
\end{figure}

\newpage
\bibliography{achemso-demo}